\documentclass[preprint,pre,aps,amsfonts,amsmath,showpacs,superscriptaddress,nofootinbib, draft]{revtex4}
\usepackage{bm}
\usepackage{feynmf}
\usepackage{amsmath}
\usepackage{graphicx}
\usepackage{amssymb,epsfig}
\usepackage{subfigure}
\usepackage{feynmf}
\begin{document}
\title{Long Range Correlated Percolation}
\author{Vesselin I. Marinov}
\email{marinov@physics.rutgers.edu}
\affiliation{Department of Physics and Astronomy,   Rutgers University, Piscataway, New Jersey 08854-8019.}
\begin{abstract}
In this note we study the field theory of dynamic isotropic percolation (DIP) with quenched randomness that has long range correlations
decaying as $r^{-a}$. We argue that the quasi static limit of this field theory describes the critical point of long range correlated percolation. 
We perform a one loop double RG expansion in $\epsilon=6-d$, d the spacial dimension, and $\delta = 4-a$ and calculate both the static exponents and
the dynamic exponent corresponding to the long range stable fixed point. The results for the static exponents as well as the region of stability for
this long range fixed point agree with the results from a previous work on the subject that used a different representation of the problem ~\cite{aweinrib}. 
For the special case $\delta = \epsilon$ we perform a two loop calculation. We confirm that the scaling relation  $\nu = \frac{2}{a}$, $\nu$ is the correlation
length critical exponent, is satisfied to two loop order. 
Simulation results for the spreading exponent in $d=3$ differ significantly from the value we obtain after Pade-Borel resummation was performed on the $\epsilon$ expansion result. This is in sharp contrast with the result of a two loop  $\epsilon$ expansion for the spreading exponent for DIP where there is a very good agreement with results
from simulations for $d \geq 2$.
\end{abstract}
\maketitle
\section{Introduction} 
For independent site percolation on $\mathbb{Z}^d$, sites are independently assigned to be open with probability
 $p$  or closed with probability $1-p$. The subject of percolation is
the study of the maximally connected sets of open sites, called
clusters,  where two sites are connected if they share a common
edge. It could be proved that there  exists a critical $p$, denoted as
$p_c$, such that for  $p > p_c$ with probability $1$ there exists an
infinite cluster of open sites and for $p < p_c$ the probability of
such an event is zero. Independent percolation is the simplest
example  of a system with a second order phase transition. The
order parameter is the probability that a given site belongs to
the infinite cluster, this is zero for $p < p_c$ and non zero for $p
> p_c$. In the case of  independent percolation it could be proved
that this order  parameter is actually continuous. For more general information
on percolation consult \cite{mathperc,introperc}  
\\
In analogy with the usual second order phase transition we can define
various critical exponents ~\cite{introperc}. The probability that
two distant sites belong to the same cluster for $p< p_c$
decays exponentially with a characteristic length (correlation length) $\xi(p)$. Near
$p_c$ the correlation length scales as $\xi(p) \sim |p-p_c|^{-\nu}$. The expected cluster size $S(p)$
for $p<p_c$ and $p$ near $p_c$ scales as $S(p) \sim |p-p_c|^{-\gamma}$. The probability that a 
given site belongs to the infinite cluster $P_{\infty}(p)$ for $p>p_c$ and $p$ near $p_c$ scales as
$P_{\infty} \sim |p-p_c|^{-\beta}$.  These exponents are not all independent, they satisfy
the hyper-scaling relation $d= 2\frac{\beta}{\nu} + \frac{\gamma}{\nu}$ ~\cite{introperc}. Another exponent which is 
of interest in this note is the spreading exponent. The shortest path $L$ 
between two points on the infinite cluster a distance $R$ apart scales as $R^{z_s}$, where 
$z_s$ is the spreading exponent ~\cite{shortest}.
\\
Percolation is equivalent to the $Q \rightarrow 1$  limit of a Q-state Potts model,
that is, the critical exponents of the Q-state Potts model in the $Q \rightarrow 1$ limit coincide with those of percolation ~\cite{cardyscale}. 
This could be seen from the partition function of the
Q-state Potts model which could be written as  a sum over "clusters" which in the $Q \rightarrow 1$ limit correspond to the 
clusters of the percolation problem. Thus, developing a field theory for the Q-state Potts model and performing an RG analysis
on this field theory allows us to 
``calculate'' some of the exponents of the independent percolation problem. Using such an approach, the critical exponents $\nu$ and $\gamma$ and thus also $\beta$, using 
the hyper scaling relation, have been computed in terms of an $\epsilon = 6-d$ expansion up to third order ~\cite{threeloop}. 
\\
The statistical properties of the clusters of independent percolation near a percolation threshold 
can also be studied using the so called General Epidemic Process(GEP). GEP is an example of an absorbing state 
phase transition. For this process a ``disease'' is spreading through a media of 
susceptible individuals. The susceptible media becomes infected with rate dependent on the density of the sick and the density
of recovered individuals. After a brief time interval the sick recover and are immune after that. The recovered individuals, sometimes referred to as debris when
GEP is used to describe the spread of fire for example, stop the spread of the disease locally. The state with zero density of sick individuals is absorbing, 
i.e. the disease can not spontaneously reappear. The statistical properties of the debris clusters that are left behind after the 
disease has been extinguished are described by independent percolation  ~\cite{janssen,cardygrass}. This description of the independent percolation problem 
allows to probe in addition to the exponents $\nu$, $\gamma$ and $\beta$ also the spreading exponent $z_s$ which is connected
to the dynamic exponent $z$ of the DIP field theory by the relation $z_s = \frac{2}{z}$ ~\cite{janssen,cardygrass}. 
\\
In this note we study a variation of the independent percolation  problem. 
We study the effective field theory of a percolation problem in which
deciding if a  given site is open or closed dependents on its
surrounding. Such problems arise naturally in statistical mechanics 
models. For example in the Ising model on $\mathbb{Z}^d$ we can declare a site for which the value of the spin is $-1$ to
be open and a site for which the value of the spin is $1$ to be closed. If we are at an infinite temperature then clearly 
this defines an independent percolation problem. If we are at a finite temperature however there will be correlations 
for the spin values at different sites, and thus we have a correlated percolation problem.
\\
If the correlations of the occupation variables for the percolation problem are 
governed by a finite correlation length, i.e. they are exponential, then the only effect of this  
correlations is to shift the critical density but they don't change the properties of the phase transition i.e. they don't change the
critical exponents ~\cite{introperc}. 
\\
A way to get different critical exponents is to have correlations
that decay at large distances as power laws, i.e. the model has an 
infinite correlation length for the occupation variables. In the example of the 
Ising model given above this will correspond to the system being at its critical temperature.
\\
In  ~\cite{weinribhalp} Weinrib and Halperin
argued that the critical exponents of the percolation transition should depend only on  the decay of 
the pair correlation $G(r)$ in such systems. In particular for $G(r) \sim r^{-a}$ the transition should be in a universality class which depends only on
$a$ and $d$, d- the dimensionality of the problem. 
Their analysis was based on considering the variance of the particle density in a region of volume $\xi^d$.
 They found that if $a<d$  these correlations are relevant if $a \nu - 2 < 0$, where $\nu$ is the correlation length critical exponent corresponding to the pure percolation
problem. Weinrib and Halperin argued that systems that satisfy the above criteria belong to a new universality class for which the percolation correlation 
length exponent is $\nu_{long} = \frac{2}{a}$.
\\
In ~\cite{aweinrib} Weinrib used the mapping of the percolation model to the $Q \rightarrow 1$ limit of a Q-state  Potts model to construct an effective field theory of the long range correlated percolation problem. The long range correlations of the percolation problem were 
mapped onto a long range correlated disorder in the couplings of the Potts model. To derive an effective field theory Weinrib had to resort to the so called ``replica trick''
and a cumulant expansion. Weinrib performed a renormalization group analysis of this field theory 
and using a double expansion in $\epsilon = 6-d$ and $\delta = 4-a$ to one loop he obtained results which agreed with the results 
from simple scaling arguments. 
\\
In this note we  take a different approach to the problem. We consider a DIP field theory for which 
the critical control parameter $\tau$ is disordered with a quenched correlated disorder that decays for large distances $r$ as $r^{-a}$. The
static properties of the clusters that are left behind after the agent has been extinguished  are described by long range 
correlated percolation. 
\\
Performing the averaging over the disorder for this dynamical model is easy, we do not need the replica trick ~\cite{gribov}. After that we 
renormalize the resulting field theory using dimensional regularization and minimal subtraction.
\\ 
The procedure of dimensional regularization and minimal subtraction results in a power series for the critical exponents 
in terms of $\epsilon$, and $\delta$ when we have double expansion. For the independent percolation problem such an expansion even only
to two loops after a Pade-Borel resummation gives good agreement for $d \geq 3$ with the values of the critical exponents obtained from 
simulation ~\cite{fucito}. For the minimal distance exponent the agreement is remarkable ~\cite{jantauber}. One might hope, even if it is not realistic,  that such an agreement might hold
for the correlated percolation problem when the power series are in terms of the two parameters $\epsilon$ and $\delta$. 
Higher than one loop double expansion however for such models seem to be difficult. In this note we consider the special 
case $\epsilon = \delta$. For such a model we perform
a two loop calculation. 
\\
The results from the two loop calculation do not agree well with simulation results, this is not very surprising. 
The power series in $\epsilon$ might not even be resummable, and even if it is, the structure of the problem is much more complicated than the one of independent 
percolation so higher loop calculation might be needed to get comparable result.
\\
Another RG approach for calculating critical exponents from field theories is the fixed dimensional renormalization, based on Parisi's ``massive'' scheme ~\cite{parisi}. Such
an approach for independent percolation for $d \geq 3$ 
up to two loops gives quite good agreement with the simulation results ~\cite{fucito}. The agreement is better than the one 
based on the $\epsilon$ expansion. Such an approach was used in ~\cite{prudnikov1} to calculate the critical exponents for $O(m)$-symmetric Ginzburg-Landau-Wilson model
 in quenched disorder with 
power law correlations for $d=3$ and $2 \leq a \leq 3$. We have tried to do similar calculation for the long range percolation problem for $d=3$  and $a$ close to $2$.
However the calculations did not reveal any fixed point other than the pure one even after Pade-Borel resumation. The details of this calculation are not presented here.   
\\
The paper is organized as follows. In section II we present the field theoretic description of the independent DIP model.
We also present a sketch of the renormalization procedure and how one  extracts the critical exponents from the so called RG equations. 
In section III we discuss our generalizations of DIP where we introduce the long range correlated quenched randomness. A first order
calculation is performed and the new fixed point corresponding to long range percolation is identified. Then 
we present the results of the two loop calculation for the special case $\epsilon = \delta$. In sec IV we present 
simulation results for the Voter model. In the 
appendix we provide some details of the calculation.  

\section{Dynamic Isotropic Percolation}
There are two ways to extract an effective field theory functional which after
an RG study  gives the critical exponents of DIP ~\cite{jantauber}. 
One  approach is to start from the master equation of the microscopic dynamics of a specific model in the DIP universality class. 
Representing this in terms of bosonic creation and annihilation operators and using coherent-states one can proceed towards
field theory ~\cite{benzoni}. A second approach is to use phenomenological arguments
and write down an effective Langevin equation obeying all the requirements and symmetries of the theory. We can map  this equation 
into a field theory functional ~\cite{janssen}. Following this general principle one arrives at the
following effective action for DIP
\begin{equation}\label{functional1}
I\left[s,\hat{s}\right] = \int d^dr dt \left\{ \hat{s} \left[\partial_t + \lambda \left(\tau -
  \bigtriangledown^2\right) + \frac{\lambda g}{2}
\left(2S - \hat{s}\right)\right]s \right\}
\end{equation}  
where $S(r,t) = \lambda \int_0^{t} dt^{'} s(r,t^{'})$ is the density of debris at
site $r$, $s(r,t)$ is  the density of infected individuals, $\tau$ is the critical control parameter, $\lambda$ is proportional to the recovery rate, and 
finally $\hat{s}$ is the response field. 
\\
The naive scaling dimensions of the fields and couplings for the DIP action are as follows
\begin{eqnarray}{\nonumber}
S \sim \hat{s} \sim \mu^{(d-2)/2} \\ {\nonumber}
s \sim \mu^{(d+2)/2} \\ {\nonumber}
g \sim \mu^{(6-d)/2} \\ {\nonumber}
\end{eqnarray}
where $\mu$ is an arbitrary scale of length an time.
We see that the upper critical dimension $d_c$ of the theory is $6$, that is for $d > 6$ the theory is 
asymptotically free.
\\
In the calculation of the Green's functions of a general field theory ultraviolet divergences arise, also infrared divergences if we are 
at the critical point. For a renormalizable field theory the divergences can be removed by absorbing them in the bare coupling constants and fields. 
For the DIP field theory we define renormalized fields and couplings as follows ~\cite{jantauber}
\begin{eqnarray}{\nonumber}
s_r = Z^{-1/2} s, \ \ \ \hat{s}_r = \hat{Z}^{-1/2} \hat{s} \\{\nonumber}
u = G_{\epsilon} g^2 \hat{Z}^{3} {Z_u}^{-1} \mu^{-\epsilon}, \ \ \
\lambda_r = (Z \hat{Z})^{1/2} \hat{Z}^{-1}  \lambda \\{\nonumber}
\tau = \hat{Z}^{-1} {Z_{\tau}} \tau_r + \tau_c, \\ {\nonumber}
\end{eqnarray}
where, $\epsilon = 6-d$, $G_{\epsilon} =
\frac{\Gamma(1+\epsilon/2)}{(4\pi)^{d/2}}$. The renormalization constants
$Z_{...} = Z_{...}(u,\mu/\Lambda,\epsilon)$ can be chosen in a
UV-renormalizable theory in such a way that 
\begin{equation}\label{green}
G^r_{N,\hat{N}}(\{r,t\}, \tau_r, u, \lambda_r, \mu) = \lim_{\Lambda
  \rightarrow \infty} Z^{-N/2} {\hat{Z}}^{-\hat{N}/2} G(\{r,t\},\tau,g,\lambda,\Lambda)
\end{equation}
and $G^r_{N,\hat{N}}$ in ~(\ref{green}) are finite and well defined, here $G(...)$ are the Green's functions of the theory. 
We have regularized the field theory in ~(\ref{green}) by introducing a high momentum cutoff $\Lambda$. For $d \leq d_c$ the
critical theory has IR singularities and for $d \geq d_c$ the theory has
UV singularities. Indeed the problematic UV and IR singularities are
linked precisely at $d=d_c$. What is important is that the determination of 
the $Z$ factors coming from the UV divergence provides 
information of the critical IR singularities and thus on the 
critical exponents ~\cite{jantauber,amit}. 
\\
In the explicit
calculation that we perform we fix $\epsilon > 0$ and take the
continuum limit $\Lambda \rightarrow \infty$ and we require that the Z
factors absorb the $\epsilon$ poles. This procedure is called minimal
subtraction. Note that in such a calculation $\tau_c$ is set to zero. By
requiring that as $\epsilon \rightarrow 0$ the theory gives finite
results we can calculate the exponents as power series in
$\epsilon$.
\\
The bare Green's functions are independent of the renormalization scale $\mu$ therefore from ~(\ref{green}) follows the Renormalization Group (RG) equation: 
\begin{equation}\label{RG}
\left[\mu \frac{\partial}{\partial \mu} + \xi \lambda_r \frac{\partial}{\partial \lambda_r} + k \tau_r \frac{\partial}{\partial \tau_r} 
+ \beta \frac{\partial}{\partial u}+ \frac{1}{2} (N \Upsilon + \hat{N} \hat{\Upsilon})\right] G^r_{N,\hat{N}}\left({r,t},\tau_r,u,\lambda_r,\mu\right)= 0
\end{equation}
where
\begin{eqnarray}{\nonumber}
\beta(u) = \mu \left.\frac{\partial u}{\partial \mu} \right|_{0} \ \ \ \Upsilon(u) = \mu \left.\frac{\partial \ln Z}{\partial \mu} \right|_{0}, \ \ \ 
\ \hat{\Upsilon}(u) = \mu \left.\frac{\partial \ln \hat{Z}}{\partial \mu} \right|_{0} \\ {\nonumber}
k(u) = \mu \left.\frac{\partial \ln \tau_r}{\partial \mu} \right|_{0}\ \ \ \xi(u) = \mu \left.\frac{\partial{\ln \lambda_r}}{{\partial \mu}} \right|_{0}. \nonumber
\\
\end{eqnarray}
The partial differential equation ~(\ref{RG}) can be solved employing the method
of characteristics. After solving the RG equation and employing 
 dimensional scaling one arrives to an asymptotic form, long distance ,
 long time of the Green's function from which the critical exponents 
could be derived ~\cite{jantauber}. The critical exponents of the percolation problem are given by
\begin{eqnarray}{\nonumber}
\frac{\gamma}{\nu} = 2 - \hat{\Upsilon}(u^*) \\ \nonumber  \frac{1}{\nu} = 2 - k(u^*) \\ \nonumber  z = 2 + \xi(u^{*}) \nonumber
\end{eqnarray}
where $u^*$ is a stable fixed point of the RG, that is $\beta(u^*) = 0 $ and $\beta^{'}(u^*) > 0$.
\\
An RG study through an $\epsilon = d-6$ expansion for DIP results in exponents
which agree with the exponents obtained from an $\epsilon$ expansion of the Q-state Potts model in the $Q \rightarrow 1$ limit. 
The reason why this is so is given in ~\cite{cardygrass}.
\\
\section{Correlated Dynamic Isotropic Percolation}
Let us introduce  a  variation of DIP in which the critical control  parameter $\tau$  that governs the strength of the infection is 
itself position dependent variable $\tau + c(r)$, with $c(r)$ some random field.
\\
If we take static Gaussian distributed disorder with correlations 
$<c(r_1),c(r_2)> \sim f \delta(r_1-r_2)$ and zero average, f the strength of the disorder, and we perform the average, 
we observe that the scaling dimension of $f$ is $4-d$. This is an irrelevant perturbation near $6$ dimensions
so we expect this kind of disorder not to change the critical behavior. 
\\
If however we assume Gaussian disorder with correlations
$<c(r_1)c(r_2)> \sim f \frac{1}{|r_1 - r_2|^a}$ and zero average, then the scaling dimension of $f$ is $4-a$ which is a relevant perturbation for $a<4$.
The explicit form of the functional is 
\begin{equation}\label{functional1}\begin{split}
I\left[s,\hat{s}\right] = & \int d^dr dt \left\{\hat{s} \left[\partial_t + \lambda \left(\tau - \bigtriangledown^2\right) + 
\frac{\lambda g}{2}\left(2S - \hat{s}\right)\right]s\right\} - \\
& \frac{\lambda^2 f}{2}\int dt_1 \int dt_2 \int d^dr_1 \int d^dr_2 \hat{s}\left(r_1,t_1\right) s\left(r_1,t_1\right) \frac{1}{|r_1 - r_2|^a} \hat{s}\left(r_2,t_2\right) s\left(r_2,t_2\right).
\end{split}
\end{equation}
If we are only interested in time independent quantities (emerging as $t
\rightarrow \infty $) it is convenient  to go to the
  quasi-static limit ~\cite{jantauber}. Taking the quasi-static limit  amounts to
  switching the fundamental field variable from the agent density $s$
  to to the final density of debris $\phi(r) := S(r,\infty) = \lambda
  \int_0^{\infty} s(r,t) dt$ that is ultimately left behind by the
  epidemic and the associated response field $\hat{\phi}(r) =
  \hat{s}(r,0)$ ~\cite{jantauber}.
\\
The structure of the action allows us directly to let 
\begin{equation} \nonumber
\hat{s}(r,t) \rightarrow \hat{\phi}(r), \ \ \ \ \phi(r) = \lambda
\int_0^{\infty} s(r,t) dt
\end{equation}
This results in the quasi-static Hamiltonian: 
\begin{equation}\label{hamiltonian1} \begin{split}
H\left[\phi,\hat{\phi}\right] = & \int d^d x \left\{\hat{\phi}\left[\tau - \bigtriangledown^2 + \frac{g}{2}(\phi
  - \hat{\phi})\right]\right\}  \\ 
&  - \frac{1}{2} f \int d^d x_1 \int d^d x_2 \phi(x_1)
  \hat{\phi}(x) \frac{1}{|x_1 - x_2|^{a}} \phi(x_2) \hat{\phi}(x_2)
\end{split}
\end{equation}
In addition to the rules coming from the Hamiltonian above we have to specify that closed propagator loops are not allowed.
\\
It is more convenient to carry the calculations in momentum space. The Fourier transform of the
interaction vertex $g(x) \sim x^{-a}$ is  $g(k) = v + w k^{a-d}$ for
small k  ~\cite{weinribhalp}. As discussed at the beginning of this section $v$ is irrelevant and will be ignored, $w>0$. We now absorb $w$ in the definition of $f$.
\\
 Using again the arbitrary inverse length scale $\mu$ by 
inspection we obtain that 
\begin{eqnarray}{\nonumber}
\hat{\phi} \sim \phi \sim \mu^{d/2} \\ {\nonumber}
g\sim \mu^{\frac{6-d}{2}} \ \ \ f \sim \mu^{4-a} \\ {\nonumber}
\end{eqnarray}
 The upper critical dimension is 6 and $f$ is relevant for $a<4$. The propagator is 
\begin{equation}
G(q) = \frac{1}{\tau + q^2}
\end{equation}
and the vertices are given by $U_1 = -U_2= g$ and $V = \frac{f}{q^{d-a}}$. To extract the divergences
we have only to calculate the one particle irreducible diagrams denoted  here by $\Gamma$.
Inspection of the naive divergence of the one particle irreducible
diagrams show that they arise only in the diagrams contributing to $\Gamma^{1,1}$,
$\Gamma^{1,2} = - \Gamma^{2,1}$ and $\Gamma^{2,2}$. Here the first
index is the number of amputated  external $\hat{\phi}$ legs and the
second is the  number of amputated $\phi$ legs. The vertex functions 
are considered as functions of external momenta and we require that $\Gamma^{1,2}(0)$, $\Gamma^{2,2}(0)$, $\Gamma^{1,1}(0)$ and $\left. \frac{d \Gamma^{1,1}}{d p^2} \right|_{p^2 = 0}$ are finite. The model is renormalizable 
by the following scheme 
\begin{eqnarray}{\nonumber}
\phi_r = \hat{Z}^{-1/2} \phi, \ \ \ \hat{\phi}_r = \hat{Z}^{-1/2} \hat{\phi},   \\ {\nonumber}
\tau_r = Z_{\tau}^{-1} \hat{Z} \tau, \\ {\nonumber}
u = G_{\epsilon} \mu^{-\epsilon} \hat{Z}^3 Z_u^{-1} g^2 ,\\ {\nonumber}
v = F_{\delta} \mu^{-\delta} \hat{Z}^2 Z_v^{-1} f.  {\nonumber}
\end{eqnarray}
where  $F_\delta
=\frac{G(1+\frac{\delta}{2})\Gamma(\frac{a}{2})}{{4\pi}^{d/2}\Gamma(\frac{d}{2})}$ and
$\delta = 4-a$. 
Evaluating to one loop order the divergent diagrams and 
using minimal subtraction and double expansion, where now we require that the $Z$ factors absorb both $\epsilon$ and $\delta$ poles,  
we obtain the following results
\begin{eqnarray}{\nonumber}
\hat{Z} = 1 + \frac{u}{6\epsilon} - \frac{4v}{6 \delta}, \\ {\nonumber}
Z_\tau = 1 + \frac{u}{\epsilon} - \frac{2v}{\delta}, \\ {\nonumber}
Z_u = 1 + \frac{4u}{\epsilon} - \frac{12v}{\delta}, \\ {\nonumber}
Z_v = 1 + \frac{2u}{\epsilon} - \frac{4v}{\delta}. {\nonumber}
\end{eqnarray}
This gives us
\begin{eqnarray}{\nonumber}
\beta(u) = u(-\epsilon + \frac{7}{2}u - 10v),\\{\nonumber}
\beta(v) = v(-\delta + \frac{5}{3}u-\frac{8}{3}v), \\{\nonumber}
 k(u,v) = \frac{5u - 8v}{6}, \\ {\nonumber}
\bar{\Upsilon}(u,v)   = -\frac{u}{6} + \frac{4v}{6}. \\ {\nonumber}
\end{eqnarray}
A nontrivial fixed point, which corresponds to long
range correlated percolation is obtained:
\begin{eqnarray}{\nonumber}
u^{*} = \frac{15 \delta - 4 \epsilon}{11}, \\ {\nonumber}
v^{*} = \frac{21 \delta - 10 \epsilon}{44}. \\ {\nonumber}
\end{eqnarray}
which finally gives us
\begin{eqnarray}
\frac{1}{\nu} =  2 - \frac{\delta}{2} \\ {\nonumber}
\frac{\gamma}{\nu} = 2 - \frac{\delta - \epsilon}{11} {\nonumber}
\end{eqnarray}
One could carry analogously to Weinrib the stability analysis for the different
fixed points and he will arrive at the same conclusion as in there.
Our model however allows us to compute the spreading exponent as well.
In order to do this calculation we have to go back to the
dynamical model. Fortunately to extract the dynamical critical
exponent we have to only calculate $\Gamma^{1,1}$, as a function of the external momentum $q$ and frequency $\omega$ From the renormalization of the derivative $\left.\frac{\partial \Gamma_{1,1,}}{\partial \omega} \right|_{q^2=\omega=0}$ we obtain 
\begin{equation}
(Z \hat{Z})^{1/2} = 1 + \frac{3u}{4 \epsilon} - \frac{2v}{\delta}
\end{equation}
and from this we conclude that 
\begin{equation}
\xi(u^*,v^*) = -\frac{7u^*}{12} + \frac{4v^*}{3} = -\frac{\epsilon}{11} - \frac{7\delta}{44},
\end{equation}
and thus 
\begin{equation}
z = 2 + \xi(u^*,v^*) = 2 -\frac{\epsilon}{11} - \frac{7\delta}{44}.
\end{equation}
We are interested in obtaining estimates for the critical exponents for long range correlated percolation for $d \geq 3$. It is quite remarkable that such estimates
for independent percolation in $d \geq 3$ coming from an $\epsilon$ expansion up to two loops agree well with simulation results ~\cite{fucito}. The agreement for the
spreading exponent is quite remarkable ~\cite{jantauber}. Although it might be unrealistic we are curious whether such an agreement might hold for the case of 
long range correlated
percolation. Unfortunately it is seems difficult to carry out a two loop double expansion in $\epsilon$, $\delta$. We note here that we have performed a fixed 
dimension renormalization for $d=3$, and $a$ close to 2, which did not result any fixed point other than the pure one even after Pade-Borel resummation was performed. 
\\
We have performed a two loop expansion for the case $\epsilon=\delta$, in this case it is just an expansion in $\epsilon$.
Such models arise naturally when the correlation are expressed in terms of the probability that a random walk starting at 
a given site will hit the origin, this probability for $d \geq 3$ is proportional to $\frac{1}{|x|^{d-2}}$, thus
$a = d-2$. Examples for such models are the Voter Model and the Massles Harmonic crystal in $d \geq 3$ ~\cite{hc_paper}. 
From the $\epsilon$ expansion we obtain
\begin{align}
Z_u  & = 1 + \frac{1}{\epsilon}(4u-6v)+ u v (\frac{69}{4 \epsilon} - \frac{39}{\epsilon^2}) + u^2(-\frac{59}{12\epsilon} + \frac{15}{\epsilon^2}) + v^2 (-\frac{145}{12 \epsilon} + \frac{22}{\epsilon^2}), \nonumber \\ \nonumber 
Z_v & = 1 + \frac{2}{\epsilon}(u-v) + u v(\frac{91}{18 \epsilon}-\frac{32}{3 \epsilon^2}) + u^2(-\frac{47}{24 \epsilon} + \frac{11}{2 \epsilon^2}) + v^2 (-\frac{91}{36 \epsilon} + \frac{10}{3 \epsilon^2}), \\ \nonumber
Z & = 1 + \frac{1}{\epsilon}(\frac{u}{6} - \frac{v}{3}) + u v (\frac{71}{144 \epsilon} - \frac{3}{4 \epsilon^2}) + u^2(-\frac{37}{432 \epsilon} + \frac{11}{36 \epsilon^2}) + v^2 ( -\frac{139}{216 \epsilon}  + \frac{5}{18 \epsilon^2}), \\ \nonumber
Z_{\tau} & = 1 + \frac{1}{\epsilon}(u-v) + u v (\frac{91}{36 \epsilon} - \frac{13}{3 \epsilon^2}) + u^2 (-\frac{47}{48 \epsilon} + \frac{9}{4 \epsilon^2}) + v^2 (-\frac{91}{72\epsilon} + \frac{7}{6 \epsilon^2}). \nonumber
\end{align}
This gives us
\begin{align}
\beta_u = (-\epsilon + \frac{7}{2} u - 5 v -\frac{671}{72} u^2 + \frac{757}{24} v u - \frac{731}{36} v^2) u \nonumber \\
\beta_v = (-\epsilon + \frac{5}{3} u - \frac{4}{3} v - \frac{193}{54} u^2 + \frac{293}{36} v u - \frac{67}{27} v^2 ) v \nonumber
\end{align}
\begin{align}
\bar{\Upsilon}(u,v) = -\frac{1}{6} u + \frac{1}{3} v + \frac{37}{216} u^2 - \frac{71}{72} v u + \frac{139}{108} v^2 \nonumber \\
k(u,v) = \frac{5}{6} u - \frac{2}{3} v - \frac{193}{108} u^2 + \frac{293}{72} uv - \frac{67}{54} v^2 \nonumber
\end{align}
We identify a long range stable fixed point:
\begin{align}
u^{*} = \epsilon + \frac{59}{88} \epsilon^2 ,\nonumber \\
v^{*} = \frac{1}{2} \epsilon + \frac{131}{176} \epsilon^2 \nonumber
\end{align}
,this results in 
\begin{align}
k(u^*,v^*) = \frac{1}{2} \epsilon ,\\ \nonumber
\bar{\Upsilon}(u^{*}, v^{*})= \frac{3}{22} \epsilon^2 \nonumber
\end{align}
From the dynamical part of the calculation we obtain
\begin{align}
(Z \bar{Z})^{\frac{1}{2}} & = 1 + \frac{1}{\epsilon}(\frac{3}{4} u - \frac{227}{384} u^2 + \frac{569}{288} v u - v - \frac{91}{72} v^2 + \frac{1}{8} u v \log(2)+\frac{5}{32} \log(2) u^2 - \frac{9}{64} \log(3) u^2) \nonumber \\
& + \frac{1}{\epsilon^2}(\frac{51}{32} u^2 - \frac{83}{24} u v + \frac{7}{6} v^2).
\end{align}
This gives us
\begin{equation}
\xi(u,v) = -\frac{7}{12} u + \frac{2}{3} v + (\frac{1747}{1728}-\frac{5}{16}\log(2) +\frac{9}{32} \log(3)) u^2 -(\frac{427}{144} + \frac{1}{4} \log(2)) u v + \frac{67}{54} v^2. \nonumber 
\end{equation}
For the dynamic exponent, and consequently the spreading exponent,
 for the long range fixed point we finally obtain
\begin{align}
z = 2 - \frac{1}{4}\epsilon + (-\frac{119}{2112} - \frac{7}{6} \log(2) + \frac{9}{32} \log(3)) \epsilon^2 \nonumber \\
z_s = \frac{2}{z} = 1 + \frac{1}{8} \epsilon + (\frac{185}{4224} + \frac{7}{32} \log(2) - \frac{9}{64} \log(3)) \epsilon^2
\end{align}
To summarize, the two loop expansion gives for the correlation length critical exponent $\nu = \frac{2}{a}$. For the ratio of critical exponents 
$\frac{\gamma}{\nu}$ we obtain to one loop $2$, compare to the result of 1.8 in ~\cite{hc_paper}, 
but there is a big correction of $-\frac{3}{22} \epsilon^2$ coming from the two loops. 
From a Pade-Borel resummation of the series for $z_s$ we obtain $z_s \approx 1.6$ for $d=3$. In the next section we report on simulation results for 
$z_s$ for the Voter model percolation problem on $\mathbb{Z}^3$ ~\cite{hc_paper}.
\section{Spreading exponent for the Voter Model percolation}
\begin{figure}
\epsfxsize=3in
\begin{center}
\leavevmode
\epsffile{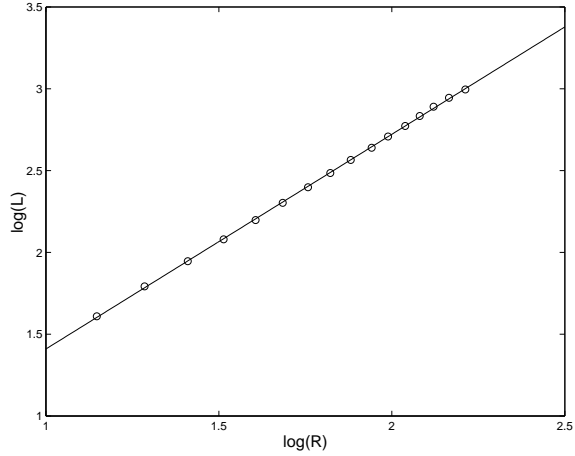}
\end{center}
\caption{Plot of $\log(R)$ versus $\log(L)$ for the 3d voter model at $p=0.1$. The slope of the
  straight line gives $z_s \approx 1.32$}
\label{fig:Fig1}
\end{figure}

To obtain the spreading exponent for independent percolation one resorts to the so called Leath algorithm. This corresponds to 
growing a cluster from a single seed ~\cite{shortest}. One could stop the growth after a certain number of ``steps'' or after the cluster hits a certain boundary, the
first is more natural. For the Voter model percolation problem this approach is not possible, we can not grow single clusters since the occupation probabilities 
are not independent. 
\\
We use the algorithm introduced in ~\cite{hc_paper} to simulate the Voter model. We pick a site and we decide that it is going to be occupied, this is our seed.
Then we run our algorithm for a cube that is centered at that site but in addition to the rules detailed in ~\cite{hc_paper} 
when a random walker hits the center we freeze it and assign all of it ancestors occupied in the percolation problem
\\
We simulate the Voter model at its critical density $p=0.1$, the results are presented in  Fig.\,\ref{fig:Fig1}. In fact not much fluctuation in the results for $z_s$ 
is observed for $p \in [0.9,0.11]$. 
\input{epsf}
We conclude that $z_s \approx 1.32$. This is smaller than the exponent of independent percolation
which is $z_s \approx 1.37$ ~\cite{shortest}. This result clearly does not agree with our $\epsilon$ expansion result. 
\section{Conclusion}
We have investigated the field theory of quenched correlated disordered DIP field theory with disorder correlations
 that decays at large distances $r$ as $r^{-a}$. We have identified a long range stable fixed point in a one loop double expansion in $\epsilon = 6-d$ and $\delta = 4-a$.  
Our results agree with the results obtained in ~\cite{aweinrib} using a different representation of the problem as well as 
different RG scheme. 
\\
For the special case $\epsilon = \delta$ we have performed an expansion to 2nd order in $\epsilon$.  
For the correlation length critical exponent we have obtained $\nu=\frac{2}{a}$, for the ratio of critical
exponents $\frac{\gamma}{\nu}$ the results is $2-\frac{3}{22} \epsilon^2$ and for the spreading exponent $z_s$ the result is 
$z_s = 1 + \frac{1}{8} \epsilon + (\frac{185}{4224} + \frac{7}{32} \log(2) - \frac{9}{64} \log(3)) \epsilon^2$. For $d=3$ after 
a Pade-Borel resummation of the series we obtained $z_s  \approx 1.6$. From a simulation of the Voter model
in 3d we have obtained $z_s \approx 1.32$.  
\\
We have also performed a fixed dimensional renormalization at $d=3$ and $a$ close to 2, the details were not reported in this note,  and we observed no fixed points other
than the pure one. It would be interesting to perform such a calculation in the case of $d=5$ and $a=3$ to see if a stable long range fixed point will appear
and how would the results compare with the result of the $\epsilon$ expansion.
\section*{Acknowledgments}
I would like to thank J. Lebowitz for discussions on the long range correlated percolation problem. The research was supported in part by 
NSF Grant DMR-044-2066 and AFOSR Grant AF-FA 9550-04-4-22910. 

\section{Appendix}
To one loop in the quasi-static limit the diagrams that contribute to the different $\Gamma 's $ are listed below.
\begin{fmffile}{temp5}
\begin{eqnarray}
\textrm{Propagator} \ \ \ 
\parbox{30mm}{\begin{fmfgraph*}(25, 18)
\fmfleft{i1} \fmfright{o2}
\fmf{fermion,label=$k$}{o2,i1}
\end{fmfgraph*}}  & \frac{1}{k^{2} + \tau}
\end{eqnarray}
\end{fmffile}

\begin{fmffile}{temp1}
\begin{eqnarray}
\textrm{Interaction vertices} \ \ \ \
\parbox{30mm}{\begin{fmfgraph*}(20, 15)
\fmfleft{i1} \fmftop{o1} \fmfbottom{o2}
\fmf{fermion}{v1,i1}
\fmf{plain}{o1,v1}
\fmf{plain}{v1,o2}
\fmfdot{v1} 
\end{fmfgraph*}}  -g \\ \nonumber
\ \ \ \ \ \parbox{30mm}{\begin{fmfgraph*}(20, 15)
\fmfleft{o1,o2} \fmfright{i1}
\fmf{plain}{v1,i1}
\fmf{fermion}{v1,o1}
\fmf{fermion}{v1,o2}
\fmfdot{v1} 
\end{fmfgraph*}} g \\ \nonumber
\ \ \ \ \ \parbox{30mm}{
\begin{fmfgraph*}(20, 15)
\fmfleft{i1,i2} \fmfright{o1,o2} 
\fmf{plain}{o1,v1}
\fmf{plain}{o2,v2}
\fmf{fermion}{v1,i1}
\fmf{fermion}{v2,i2}
\fmf{photon,label=$k$}{v1,v2}
\fmfdot{v1,v2}
\end{fmfgraph*}} \frac{f}{k^{d-a}} \nonumber
\end{eqnarray}
\end{fmffile}

\begin{fmffile}{temp2}
\begin{eqnarray}
\Gamma^{1,1} \ \ 
\parbox{30mm}{\begin{fmfgraph*}(25,20)
\fmfleft{i1} \fmfright{o1} 
\fmf{fermion}{v1,i1}
\fmf{plain}{o1,v2}
\fmf{plain,left}{v1,v2,v1}
\fmfdot{v1,v2}
\end{fmfgraph*}
}
\parbox{30mm}{
\begin{fmfgraph*}(25,20)
\fmfleft{i1} \fmfright{o1} 
\fmf{fermion}{v1,i1}
\fmf{plain}{o1,v2}
\fmf{photon,left}{v2,v1}
\fmf{plain,right}{v2,v1}
\fmfdot{v1,v2}
\end{fmfgraph*}
} \nonumber
\end{eqnarray}
\end{fmffile}
\begin{fmffile}{temp3}
\begin{eqnarray}
\Gamma^{1,2} \ \ 
\parbox{30mm}{
\begin{fmfgraph*}(25,18)
\fmfleft{i1} \fmftop{o1} \fmfbottom{o2} 
\fmf{fermion}{v3,i1}
\fmf{plain}{o1,v1} 
\fmf{plain}{v1,v3}
\fmf{plain}{o2,v2} 
\fmf{plain}{v2,v3}
\fmffreeze
\fmf{plain}{v2,v1}
\fmfdot{v1,v2,v3}
\end{fmfgraph*}
}
\parbox{30mm}{
\begin{fmfgraph*}(25,18)
\fmfleft{i1} \fmftop{o1} \fmfbottom{o2} 
\fmf{fermion}{v3,i1}
\fmf{plain}{o1,v1} 
\fmf{plain}{v1,v3}
\fmf{plain}{o2,v2} 
\fmf{plain}{v2,v3}
\fmffreeze
\fmf{photon}{v2,v1}
\fmfdot{v1,v2,v3}
\end{fmfgraph*}
} \nonumber
\end{eqnarray}
\end{fmffile}

\begin{fmffile}{temp4}                 
\begin{eqnarray}
\Gamma^{2,2} \ \ 
\parbox{30mm}{
\begin{fmfgraph*}(25,18)
\fmfleft{i1,i2} \fmfright{o1,o2} 
\fmf{fermion}{v1,i1}
\fmf{plain}{o1,v1}
\fmf{photon}{v1,v2}
\fmf{fermion}{v3,i2} 
\fmf{plain}{o2,v4}
\fmf{plain}{v4,v2} 
\fmf{plain}{v2,v3}
\fmf{plain}{v4,v3}
\fmfdot{v1,v2,v3,v4}
\end{fmfgraph*}
}
\parbox{30mm}{
\begin{fmfgraph*}(25,18)
\fmfleft{i1,i2} \fmfright{o1,o2} 
\fmf{fermion}{v1,i1}
\fmf{plain}{o1,v1}
\fmf{photon}{v1,v2}
\fmf{fermion}{v3,i2} 
\fmf{plain}{o2,v4}
\fmf{plain}{v4,v2} 
\fmf{plain}{v2,v3}
\fmf{photon}{v4,v3}
\fmfdot{v1,v2,v3,v4}
\end{fmfgraph*}
} \nonumber
\end{eqnarray}
\end{fmffile}
We evaluate those using dimensional regularization. 
\\
For the dynamical theory the propagator and vertices are
 \begin{fmffile}{temp7}
\begin{eqnarray}
\textrm{Propagator}  \ \ \ \ \ \ 
\parbox{30mm}{\begin{fmfgraph*}(25, 18)
\fmfleft{i1} \fmfright{o2}
\fmflabel{$t$}{i1}
\fmflabel{$0$}{o2}
\fmf{fermion,label=$k$}{o2,i1}
\end{fmfgraph*}}  & \theta(t)\exp(-\lambda(\tau+k^2)t)
\end{eqnarray}
\end{fmffile}

\begin{fmffile}{temp6} 
\begin{eqnarray} \nonumber
\textrm{Interaction vertices} \ \ \ \ \ \ \ \ \ \ \
\parbox{30mm}{\begin{fmfgraph*}(20, 15) 
\fmfleft{i1} \fmftop{o1} \fmfbottom{o2}
\fmflabel{$t^{'}$}{o1}
\fmf{fermion}{v1,i1}
\fmf{dashes}{o1,v1}
\fmf{plain}{v1,o2}
\fmfiv{l=t,l.a=-180,l.d=0.01*w}{c}
\end{fmfgraph*}}  -\lambda^2g\theta(t-t^{'})  \\ \nonumber
\parbox{30mm}{\begin{fmfgraph*}(20, 15)
\fmfleft{o1,o2} \fmfright{i1}
\fmf{plain}{v1,i1}
\fmf{fermion}{v1,o1}
\fmf{fermion}{v1,o2}
\fmfdot{v1} 
\end{fmfgraph*}} \lambda g \\ \nonumber
\parbox{30mm}{
\begin{fmfgraph*}(20, 15)
\fmfleft{i1,i2} \fmfright{o1,o2} 
\fmf{plain}{o1,v1}
\fmf{plain}{o2,v2}
\fmf{fermion}{v1,i1}
\fmf{fermion}{v2,i2}
\fmf{photon,label=$k$}{v1,v2}
\fmfdot{v1,v2}
\end{fmfgraph*}} \frac{\lambda^2 f}{k^{d-a}} \nonumber
\end{eqnarray}
\end{fmffile}
Notice the appearance of a time delocalized vertex ~\cite{janssen}. 

For the two loop calculation we consider all topologically different diagrams that can be obtained with our
vertices and propagator, we discard all diagrams which contain closed propagator loops.
\\
The values of the diagrams that appear could be represented as sums of three types of integrals,
or their derivatives with respect to the parameters, $a$,$b$ and $c$.
\begin{align} \nonumber
I_1(a,b,c) &= \int \frac{d^dk_1 d^dk_2}{(a+k_1^2)(b+k_2^2)(c+(k_1+k_2)^2)} \\ \nonumber
I_2(a,b,c)&= \int \frac{d^dk_1 d^dk_2}{(a+k_1^2)(b+k_2^2)(c+ k_1^2 + (k_1+k_2)^2)} \\ \nonumber
I_3(a,b,c)&= \int \frac{d^dk_1 d^dk_2}{(a+k_1^2)(b+k_2^2)(c+ k_1^2 + k_2^2 + (k_1+k_2)^2)} \nonumber
\end{align}
Only integrals of type $I_1$ appear in the calculation of the quasi static limit, while all types of integrals appear in the calculation of the dynamic exponent. 
The integral $I_1$ and its derivatives are evaluated at the point $a=b=c=1$, integral $I_2$ and its derivatives are evaluated at the point $a=b=1,c=2$ and 
integral $I_3$ and its derivatives are evaluated at the point $a=b=1,c=3$.
The dimensional regularized form of the integrals $I_1$ and $I_3$ can be found in the literature ~\cite{integrals}.
\begin{align}
I_1(a,b,c) & = \frac{1}{6 \epsilon}G_{\epsilon}^2(\left(\frac{1}{\epsilon} + \frac{25}{12}\right)\left(a^{3-\epsilon} + b^{3-\epsilon} + c^{3-\epsilon}\right) \\ \nonumber
& -\left(\frac{3}{\epsilon} + \frac{21}{4}\right)\left(a^{2-\epsilon}(b+c)+b^{2-\epsilon}(a+c)+c^{2-\epsilon}(a+b)\right) -3abc) \\ \nonumber
I_3(a,b,c) & =  G_{\epsilon}^2 ( \frac{1}{\epsilon^2}(\frac{5}{24}(a^{3-\epsilon}+b^{3-\epsilon})-\frac{1}{4}(a^{2-\epsilon}b+b^{2-\epsilon}a)-\frac{1}{8}(a^{2-\epsilon} + b^{2-\epsilon})c) \\ \nonumber
& + \frac{1}{\epsilon}((\frac{143}{288}-\frac{9}{16} \log(\frac{4}{3}))(a^3+b^3)+(\frac{1}{12}\log(\frac{4}{3})-\frac{1}{36})c^3 \\ \nonumber
& -(\frac{1}{16} + \frac{9}{8} \log(\frac{4}{3}))(a^2b+b^2a)+(\frac{1}{8}-\frac{1}{2}\log(\frac{4}{3}))c^2(a+b) \\ \nonumber
& +(\frac{15}{16} \log(\frac{4}{3}))c(a^2+b^2)+(\frac{3}{2} \log(\frac{4}{3})-\frac{1}{2})abc)) \nonumber
\end{align}
For our calculation we only need the first, or higher, derivative of $I_2(a,b,c)$ with respect to c for that we have obtained:
\begin{align}
 -\frac{\partial I_2(a,b,c)}{\partial c} &= G_{\epsilon}^2 (\frac{1}{\epsilon^2}(\frac{1}{8} a^2 + \frac{1}{2} b^2) + \frac{1}{\epsilon}(-\frac{1}{8} \log(2) a^2 \\ \nonumber
&+ \frac{9}{8} b^2 + \log(2) bc-\frac{1}{8}a^2\log(a) - \log(2)b^2 -\frac{1}{4} \log(2) c^2 \\ \nonumber
&-\frac{1}{2} b^2 \log(b) + \frac{9}{32} a^2 + \frac{1}{4} c^2 - \frac{1}{2} \log(2) ab + \frac{1}{2} \log(2) ac + \frac{1}{4} ac - \frac{1}{2} bc) )
\end{align}

\end{document}